\documentclass[12pt]{article}
\usepackage[margin=1.25in]{geometry}

\usepackage{natbib}
\usepackage{amsmath, bm}
\usepackage{amssymb}
\usepackage{graphicx}
\usepackage{multicol}
\usepackage{float}
\usepackage{hyperref}
\usepackage{color}
\usepackage{rotating}
\usepackage{subfig}
\usepackage{enumitem}
\usepackage{mathtools}
\usepackage{setspace}
\usepackage{verbatim}
\usepackage{fancyvrb}
\RequirePackage{etoolbox}
\RequirePackage{xcolor}
\definecolor{cb-blue}{RGB}{0, 109, 219}
\definecolor{cb-rose}{RGB}{255, 109, 182}

\DeclarePairedDelimiter\abs{\lvert}{\rvert}%
\makeatletter
\let\oldabs\abs
\def\abs{\@ifstar{\oldabs}{\oldabs*}}

\makeatletter
\def\namedlabel#1#2{\begingroup
    #2%
    \def\@currentlabel{#2}%
    \phantomsection\label{#1}\endgroup
}
\makeatother

\begin{document} 

\title{\large \bf Privacy Protection for Youth Risk Behavior Using Bayesian Data Synthesis: A Case Study to the YRBS}
\author{Yixiao Cao\footnote{Vassar College, 124 Raymond Ave, Poughkeepsie, NY 12604, ycao@vassar.edu}  $\,$ and Jingchen Hu\footnote{Vassar College, 124 Raymond Ave, Poughkeepsie, NY 12604, jihu@vassar.edu}}
\date{\today}
\maketitle

\begin{abstract}
 The large number of publicly available survey datasets of wide variety, albeit useful, raise respondent-level privacy concerns. The synthetic data approach to data privacy and confidentiality has been shown useful in terms of privacy protection and utility preservation. This paper aims at illustrating how synthetic data can facilitate the dissemination of highly sensitive information about youth risk behavior by presenting a case study of synthetic data for a sample of the Youth Risk Behavior Survey (YRBS). Given the categorical nature of almost all variables in YRBS, the Dirichlet Process mixture of products of multinomials (DPMPM) synthesizer is adopted to partially synthesize the YRBS sample. Detailed evaluations of utility and disclosure risks demonstrate that the generated synthetic data are able to significantly reduce the disclosure risks compared to the confidential YRSB sample while maintaining a high level of utility. 
\end{abstract}

\section{Introduction}

Respondent-level data, also known as microdata, have been widely available in public databases and are essential for students, researchers, and corporate analysts to understand a variety of research questions. Such data are typically collected through surveys and censuses, after which the data holders disseminate these data to the public. Any data dissemination needs to follow legal and ethical guidelines, which are in place to protect the privacy and confidentiality of the respondents.


The privacy and confidentiality concerns of releasing microdata could impact different communities to various extents. Not surprisingly, youth is one of the most vulnerable groups when faced with privacy intrusions. According to the Future of Privacy Forum (FPF)\footnote{\urlstyle{same}\url{https://fpf.org/blog/future-of-privacy-forum-releases-new-youth-privacy-and-data-protection-infographic/}}, consequences for youth data disclosure are severe as they are more likely to encounter predators or become victims of bullying and harassment. Less visible risks include commercial exploitation through profiling and behavioral advertising \citep{park_vance_2021}. Policy-makers and legislators across the globe have striven to sheild the privacy of data collected from youth; examples include the Children's Online Privacy Protection Act (COPPA)\footnote{\urlstyle{same}\url{https://www.ftc.gov/legal-library/browse/rules/childrens-online-privacy-protection-rule-coppa}} in the United States and the General Data Protection Regulation (GDPR)\footnote{\urlstyle{same}\url{https://gdpr-info.eu/}} in the EU. 

In this paper, we provide a case study of protecting youth data using the synthetic data approach. Our case study focuses on a particularly high-risk and vulnerable database involving youth, the Youth Risk Behavior Survey (YRBS) in the United States. \par 


\subsection{The YRBS Data}


The Youth Risk Behavior Surveillance System (YRBSS) was developed in 1990 by the U.S. Centers for Disease Control and Prevention (CDC) to monitor health behaviors that contribute markedly to leading causes of death, disability, and social problems among youth in the United States. The YRBS is the primary mechanism through which the institution collects data. The YRBS data have been extensively used by researchers and social activists to study youth behavior as well as to promote change. For example, \citet{reising_cygan_2019} provides a guide for school nurses to implement the YRBS, access results, and apply findings in their school communities, and \citet{underwood_brener_halpern-felsher_2020} discusses the strengths and weaknesses of the YRBS in tracking adolescent health behavior. \par

Given the nature of the questions asked in the YRBS, the responses are often sensitive: individuals are asked about their use of substance, sexual behavior, mental health conditions, among other things. It is important to stress that since the respondents are predominantly minors, disclosure of these sensitive information can cause legal, financial, and social consequences to the targeted minor, leading to imprisonment, detention, violence, bullying, or other types of physical and mental harms.  
In addition, the risk of disclosure would discourage YRBS respondents from answering these survey questions truthfully, as they might be concerned about their privacy and the risk of being identified, resulting in potential reductions of the survey quality. Given these reasons, it is undoubtedly important to protect the privacy of the YRBS data before their public release. The publicly available YRBS data has undergone some primary privacy protections during the data collection stage, mainly through administering the surveys anonymously and voluntarily among the students. To the best of our knowledge, little has been done to protect privacy and confidentiality at the data processing stage according to the methodology guide of the YRBS\footnote{For a detailed methodology guide of the YRBS, see: \urlstyle{same}\url{https://www.cdc.gov/mmwr/pdf/rr/rr6201.pdf}.}. For the purpose of the case study, we download a sample from the publicly available source and treat it as the confidential data.\par




We retrieve the YRBS data from the YRBSS section of the CDC website. The district-level dataset of high school students contains 504,249 observations from multiple districts across the U.S. from 1991 to 2019. For illustration purpose, we primarily focus on the 2019 survey in New York City and Chicago. 

The retrieved YRBS data contain variables such as respondent ID and sample site, demographic variables such as age, sex, and race, body mass index (BMI) variables, sexual minority variables, and the 2019 questionnaire and supplemental variables. We primarily focus on the variables that might present the biggest privacy concerns. Our selected variables are summarized in Table \ref{tab:varlist}.

\begin{table}[t]
\centering
\caption{YRBS categorical variable names, levels, and sensitive status.}
\label{tab:varlist} 
\resizebox{0.9\textwidth}{!}{
  \begin{tabular}{llc}
  \hline
  Variable name & Characteristics & Sensitive \\
  \hline
  City & New York City/Chicago & No\\
  Age &  12 to 18 years old; seven levels & No \\
  Sex & Male/female & No \\
  Grade & 9th to 12th grade; four levels & No \\
  Race & Seven categories & No \\
  Obesity indicator & Yes/No (1 and 2)& Yes\\
  Sexuality & Four categories & Yes \\
  Ever experienced sexual violence & Yes/No (1 and 2) & Yes \\
  Current tobacco use* & Yes/No (1 and 2) & Yes \\
  Current alcohol use & Yes/No (1 and 2) & Yes \\
  Current marijuana use & Yes/No (1 and 2) & Yes \\
  Ever illicit drug use** & Yes/No (1 and 2) & Yes \\
  Ever sexual intercourse & Yes/No (1 and 2) & Yes \\
  \hline
  *smoke cigarettes, electronic vapor, or cigars \\
  **ever used cocaine, heroin, or methamphetamine
  \end{tabular}}
\end{table}



Variables related to tobacco use and illicit drug use are created by combining some sub-categories, while the other variables remain the same format as in the YRBS. After removing missing values, we arrive at a sample containing $n = 5,949$ observations with $13$ variables. All variables are categorical. We deem variables related to substance use, sex, and violence sensitive and therefore to be synthesized for protection (all variables with ``Yes" in the ``Sensitive" column in Table \ref{tab:varlist}).\par

\subsection{The Synthetic Data Approach}
One approach to providing privacy protection for microdata is to generate synthetic data to be released in place of the confidential data \citep{rubin_1993, little_1993}. 
Since its first proposal almost 3 decades ago, the field has witnessed a great amount of research efforts to develop theories and models for releasing synthetic microdata. Given the fact that a subset of our YRBS variables are deemed sensitive, we follow the partially synthetic data approach, where only sensitive variables are replaced by synthetic values while non-sensitive variables remain unchanged \citep{little_1993}. One way to generate partially synthetic data is to first fit Bayesian models with the confidential data to estimate the posterior distributions. One then simulates synthetic values for the sensitive variables given the posterior predictive distributions. With carefully designed Bayesian models, the resulting synthetic data could preserve important statistical characteristics of the confidential data such as means, variances, and joint probability distributions. Moreover, they can protect the privacy of the confidential data by reducing the disclosure risks of the respondents, such as preventing intruders from identifying or inferring the values of sensitive variables for a particular individual. For a detailed overview of synthetic data, see \citet{drechsler_2011}. \par

Given the categorical nature of all of our YRBS variables, we adopt the DPMPM synthesizer, which has been shown effective for survey \citep{hu_reiter_wang_2014} and administrative \citep{drechsler_hu_2021} data. The DPMPM synthesizer is implemented by the \texttt{NPBayesImputeCat} R package \citep{npbayesimputecat} to generate five partially synthetic YRBS datasets. 
We next extensively evaluate the utility and disclosure risks of the resulting synthetic data and conclude their effectiveness of providing useful public release of the YRBS sample with sufficient privacy protection. \par

The remainder of this paper is organized as follows: Section \ref{sec:DPMPM} describes our adopted DPMPM synthesizer and our implementation details. Section \ref{sec:utility} evaluates the utility of the synthetic data while Section \ref{sec:risk} evaluates the disclosure risks. we conclude the paper with a some discussions and remarks in Section \ref{sec:conclusion}.\par



\section{The DPMPM Synthesis Model and Implementation}
\label{sec:DPMPM}
The aforementioned categorical nature of our YRBS data prompts us to adopt the Dirichlet Process mixture of products of multinomials (DPMPM) synthesis model. Compared to popular sequential synthesis models such as the classification and regression tree (CART) proposed by \citet{CART2005} where each sensitive variable is synthesized from a univariate model, the DPMPM takes the joint modeling approach by specifying a joint multivariate distribution of categorical variables. Works such as \citet{hu_reiter_wang_2014} and  \citet{drechsler_hu_2021} have demonstrated its effectiveness in synthesizing survey and administrative data.


Suppose we have the sample $\mathbf{Y}$ with $n$ observations and $r$ unordered categorical variables, where each record $i$ is denoted as $\mathbf{Y}_i = (Y_{i1},...Y_{ir})$.
The DPMPM synthesis model assumes that each $\mathbf{Y}_i$ belongs to one of $K$ underlying latent classes. The latent classification is, by definition, unobserved and therefore requires estimation. Given the latent class assignment $z_i$ of record $\mathbf{Y}_i$, each categorical variable $j$, i.e., $Y_{ij}$, independently follows a multinomial distribution where $d_j$ is the number of categories in variable $j$ ($j = 1, ..., r$). Mathematically:
\begin{align}
    Y_{ij} \mid z_i,\theta \stackrel{ind}{\sim} \text{Multinomial}(\theta_{z_i1}^{(j)}, ...,\theta_{z_id_j}^{(j)};1) \;\; \forall i, j, \label{eq:Y}\\
    z_i \mid \bm{\pi} \sim \text{Multinomial}(\pi_1,...,\pi_K; 1) \;\; \forall i, \label{eq:z}
\end{align}
where $\bm{\pi}$ is the probability vector of the latent class assignment and $\bm{\theta}^{(j)}_{k}$ is the probability vector of the categories of variable $j$ for latent class $k$. One way to estimate the model parameters is to use the truncated stick-breaking representation of the Dirichlet process priors following \citet{sethuraman_1994}. 

We implement the Markov chain Monte Carlo (MCMC) estimation process using the \texttt{NPBayesImputeCat} R package \citep{npbayesimputecat}. It uses a blocked Gibbs sampler to estimate the joint posterior distribution and provides posterior draws of all model parameters, from which synthetic data can be generated. We report the utility and disclosure results in the next sections based on $m = 5$ simulated synthetic datasets as the results are not sensitive to $m \geq 5$. \citet{RJ-2021-080} presents detailed instructions of using the \texttt{NPBayesImputeCat} R package for data synthesis. We include our R script below for interested readers.

\begin{small}
\begin{verbatim}
YRBS_syn <- NPBayesImputeCat::DPMPM_nozeros_syn(
  X = YRBS_data, 
  dj = dj, 
  nrun = 10000, 
  burn = 5000,
  thin = 10, 
  K = 80, 
  aalpha = 0.25, 
  balpha = 0.25,
  m = 5, 
  vars = c("obesity","sexuality","sexual_violence","tobacco",
           "alcohol","marijuana","drug","sexual_contact"),
  seed = 221, 
  silent = TRUE)
\end{verbatim}
\end{small}

\section{Utility Evaluation and Results}
\label{sec:utility}
For synthetic data to be released, a key criterion is being useful, i.e., they should preserve characteristics of the confidential data. 
Two types of utility are typically considered in the literature: global utility and analysis-specific utility. The former evaluates the closeness between the confidential and synthetic data distributions, while the latter evaluates whether synthetic data users can obtain inferences on the synthetic data that are similar to those obtained from the confidential data \citep{woo_reiter_oganian_karr_2009, snoke_raab_nowok_dibben_slavkovic_2018}. We consider a few metrics of each in our utility evaluation of the resulting synthetic YRBS data.


\subsection{Global Utility}
We evaluate the global utility of the synthetic data through propensity scores (pMSE) and the distribution of differences in relative frequencies for cross-tabulations. As the results show, both measurements indicate that our synthetic data preserve a high level of global utility.

\subsubsection{Propensity scores (pMSE)}
Propensity score measures the probability for individuals in a dataset being assigned to a specific treatment group given their information on other variables. It is commonly used in causal inference to reduce bias from confounding variables when estimating the effect of an intervention in an observational study. \citet{woo_reiter_oganian_karr_2009} first proposed using it for measuring global utility in the case of synthetic data and and the methodology is further expanded by \citet{snoke_raab_nowok_dibben_slavkovic_2018}. In this context, the treatment is whether the data are synthesized, and the variables used to estimate the probability are all variables in the datasets.\par

The evaluation takes place for each of the $m$ synthetic datasets. First, we combine the confidential and the synthetic datasets into one. Assume the confidential dataset has $n_c$ records and the synthetic dataset has $n_s$ records, we arrive at a concatenated dataset with dimension $(n_c+n_s)$-by-$r$, where $r$ is the number of variables. Next, we create an additional binary variable $S$ for each record indicating whether it belongs to the synthetic or confidential data, i.e., $S_i = 1$ if synthetic and $S_i = 0$ if confidential. With this setup, for each record, we can use the $r$ variables to predict the probability of $S_i$ taking value 1, which is the estimated propensity score, denoted as $\hat{p}_i$. In our case study, a logistic regression is used for the prediction of $\hat{p}_i$ \par

The propensity score mean-squared error, known as $\text{pMSE}$, is computed as
$\text{pMSE} = 1 / (n_c+n_s)\sum_{i=1}^{n_c+n_s}(\hat{p_i}-c)^2$
where $c$ is the proportion of units with synthetic data, i.e., $c = n_s / (n_s+n_c)$. In our case, for each of our $m = 5$ partially synthetic datasets, we compute the pMSE where $n = n_c = n_s$ and $c = 0.5$. As can be seen from its mathematical form, the pMSE is a measurement of how well a model can differentiate between the confidential and the synthetic dataset given all variables. It measures the deviation of the predicted probability from $c = 1/2$, i.e., how much more certain the model is at telling the difference between two datasets than a random guess. Therefore, the smaller the pMSE score, the poorer the model is at distinguishing the two datasets, thus the higher the utility. \par

The average pMSE score computed from our $m = 5$ synethtic datasets is $\mathbf{0.009}$, indicating high global utility of our synethetic data. However, a major limitation of the pMSE measurement is that it is model-dependent, i.e., the pMSE result depends on the model used for distinguishing the two datasets. The logistic regression is presumably a relatively ``weak" model, and more complex algorithm might potentially do a better job in separating the two datasets. See \citet{snoke_raab_nowok_dibben_slavkovic_2018} for further discussion.

\subsubsection{Absolute deviation and differences in relative frequency }
For categorical data, \citet{drechsler_hu_2021} considered the distributions of differences in relative frequencies between the confidential data and synthetic data for various tabulations as a measurement of global utility. For any cross-tabulation of categorical variables, we compute the relative frequency of each cell entry as $c_{jv}^{(t)}$ and $s_{jv}^{(t)}$ ($c$ for confidential and $s$ for synthetic) for $t$th cross-tabulation with $j$th variable and $v$th category. The relative difference is then computed as 
\begin{equation}
d^{(t)}_{jv} = \frac{s_{jv}^{(t)}-c_{jv}^{(t)}}{c_{jv}^{(t)}},
\end{equation}
obtaining a matrix $\bm{d}^{(t)}$ for each cross-tabulation $t$. The distribution of $\bm{d}^{(t)}$ for one-way cross-tabulations centers at 0 and ranges from -2 to 2 while that for two-way cross-tabulations centers at 0 ranging from -5 and 5. Plots are included in Appendix \ref{density plots}. 

We further consider $|s_{jv}^{(t)}-c_{jv}^{(t)}|$ as the absolute deviation between the two datasets. The smaller the absolute deviation, the closer the two datasets, indicating high global utility. The average absolute deviation for one-way, two-way, and three-way cross-tabulations are $\mathbf{0.005}$, $\mathbf{0.006}$, and $\mathbf{0.005}$, respectively, suggesting high global utility.

As with the pMSE metric, both the relative frequency difference and the absolute deviation metrics suggest a high level of global utility of our synthetic YRBS datasets.

\subsection{Analysis-specific Utility}
The analysis-specific utility measures are tailored to the analyses expected to be performed on the synthetic data. The expectation is that a data analyst would obtain similar inferences from the synthetic and the confidential data. To evaluate our synthetic YRBS data, two metrics of analysis-specific utility are considered: inference for a point estimate and inference for regression coefficients.\par

\subsubsection{Inference for a point estimate}
Since the synthesized variables are all categorical, important point estimates are proportions. We believe the proportion of heterosexual students is a highly useful quantity to report, which has a point estimate of $\hat{p}_c = 0.817$ and a 95\% confidence interval of (0.807, 0.827) in the confidential data. 


The point estimate and 95\% confidence interval for the $m = 5$ synthetic YRBS datasets can be obtained by using combining rules for partially synthetic data \citep{drechsler_2011}. Specifically, the point estimate is the mean of the point estimate from the $m = 5$ synthetic datasets $\bar{q}_m$, and the variance estimate is expressed as $T_p = b_m / m + \bar{v}_m$ 
where $b_m$ is the cross-sample variance of the proportions $p^{(l)}$ for each sample $l$ and $\bar{v}_m$ is the mean of the $m$ sample variances. The point estimate for the $m = 5$ synthetic datasets is $\hat{p}_s = 0.815$ with a $95\%$ confidence interval of $(0.801, 0.830)$. 







To evaluate the closeness between the two confidence intervals, we compute the interval overlap metric described in \citet{drechsler_reiter_2009}:
$I = (U_i-L_i)/2(U_c-L_c)+ (U_i-L_i)/2(U_s-L_s)$
where $L_s, L_c$ denote the lower CI bound of the synthetic and confidential datasets, $U_s, U_c$ denote the upper CI bound of the two datasets, and $L_i = \max(L_s, L_c)$, $U_i = \min(U_s, U_c)$. The highest possible value of $I$ is 1 and our synthetic data yield an overlap of $\mathbf{0.845}$, indicating high utility for this particular point estimate. We compute the same metric for the proportions of other synthesized variables, including tobacco use, alcohol, marijuana, sexual contact, and drug. All show an interval overlap above 0.885 with the exception for marijuana of 0.770 and drug of 0.310. The low overlap of drug is due to a very small fraction of drug users. \par



\subsubsection{Inference for regression coefficients}
\begin{figure}[h!]
    \centering
    \caption{Point estimates, confidence intervals, and interval overlaps for selected regression coefficients in a logistic regression analysis.}
    \includegraphics[width = \textwidth]{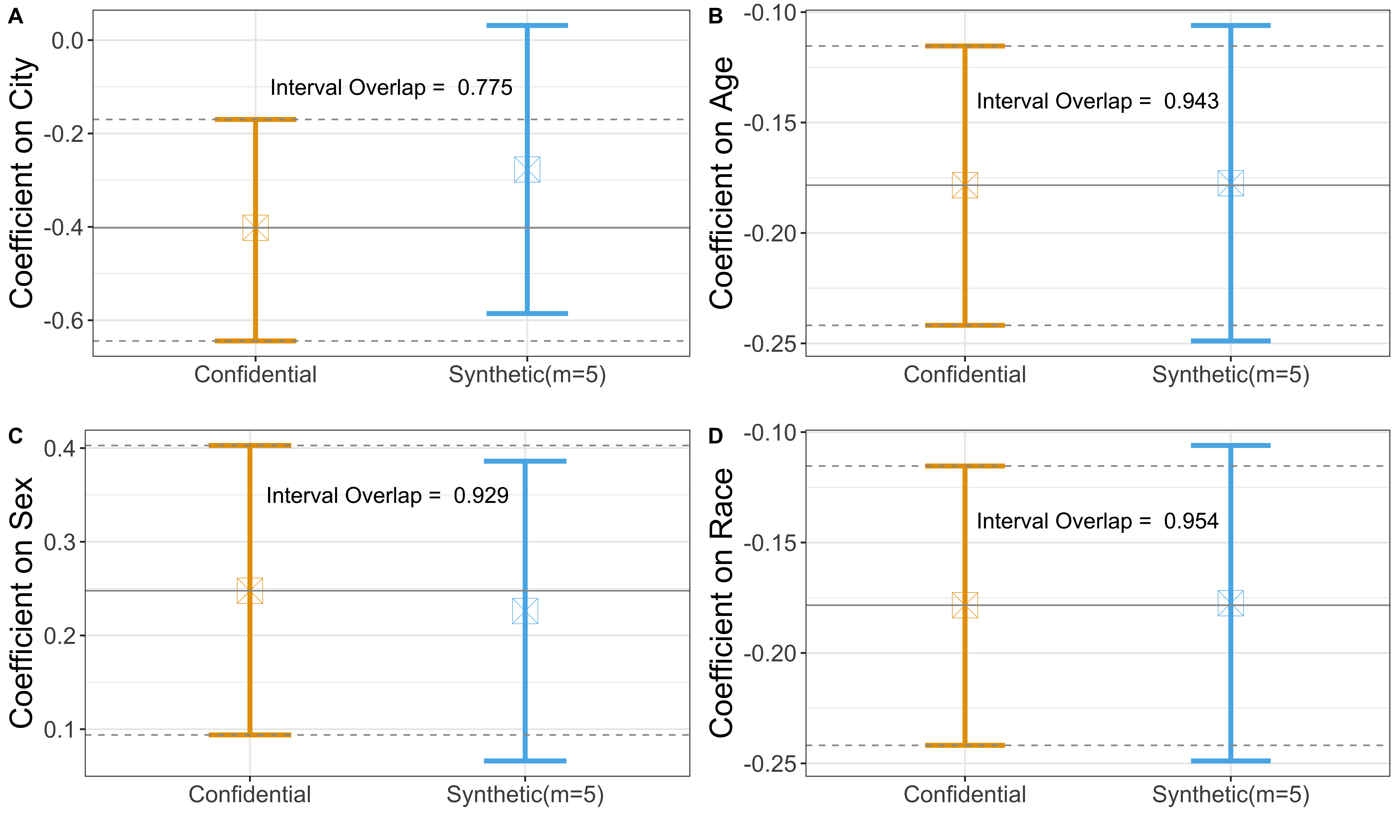}
    \label{fig:reg_coef}
\end{figure}

Similar to the inference for a point estimate, we can imagine the data analyst is conducting regression analysis, using some variables to predict the others. For example, one might want to use city, age, sex, and race to predict tobacco use with a logistic regression model. The point estimates and 95\% confidence intervals for selected regression coefficients from the confidential data and the synthetic data are obtained and visualized in Figure \ref{fig:reg_coef}. As before, we use appropriate combining rules for the synthetic data and include interval overlap metrics in the plots.

Evidently, the interval overlaps for all considered regression coefficients are extremely high with the exception for the coefficient of city, indicating an overall high level of analysis-specific utility. We run similar regressions on other combinations of variables and obtain similar results.



In summary, our partially synthetic YRBS data preserve a high level of utility both in terms of global and analysis-specifc utility considering a series of metrics. We now turn to the evaluation of their disclosure risks.

\section{Disclosure Risk Evaluation and Results}
\label{sec:risk}

The primary objective of releasing synthetic data in place of confidential data is to provide privacy and confidentiality protection. Therefore, an important aspect of synthetic data evaluation is to measure the extent to which synthetic data can reduce disclosure risks. Only when the disclosure risks of generated synthetic data are acceptable by the data disseminators can synthetic data be released to the public.

We consider two types of disclosures: identification disclosure and attribute disclosure. As the names suggest, identification disclosure is when the intruder correctly identifies records of interest, and attribute disclosure is when the intruder correctly infers the true confidential values of the synthetic variables \citep{hu_2019}. 

\subsection{Identification Disclosure}





We consider two approaches to evaluate identification disclosure risk: the matching-based approach and the record linkage approach. Both approaches show that our synthetic YRBS have significantly reduced the identification disclosure risks compared to the confidential YRBS.

\subsubsection{Matching-based approach}

In the matching-based approach, we assume the intruder possesses some knowledge for a confidential record $i$ and tries to identify the individual associated with this record $i$ in the released synthetic data \citep{reiter_mitra_2009}. Specifically, we consider specific scenarios that an intruder might encounter and quantify the corresponding disclosure risks using the following three metrics: 1) expect match risk, the expected number of correct identity matches in the released synthetic data; 2) true match rate, the percentage of true and unique matches; and 3) false match rate, the percentage of unique matches that are false matches. Appendix \ref{key_quantities} includes detailed definitions of these three metrics.


In our synthetic YRBS, we assume the un-synthesized city, age, sex, grade, and race are variables available to the potential intruder. We compute the aforementioned three metrics for both the synthetic and the confidential data to evaluate the reduction of disclosure risks. For the synthetic data, we take the average of the metrics over the $m = 5$ synthetic datasets. The \texttt{IdentificationRiskCalculation} R package is used for these implementations \citep{hornby_and_hu}. Results are summarized in Table \ref{disclosure_risk1}.
\begin{table}[h!]
    \caption{Identification risk summaries based on the matching-based approach.}\label{disclosure_risk1}
\begin{center}
\begin{tabular}{lrr  lrr}
     & Confidential & Synthetic & & Confidential & Synthetic\\
     \hline 
     Expected risk & 2234 & 186 & \hspace{0.5cm}False match rate & 0 & 0.891\\ 
     True match rate & 0.257  &  0.012 & \hspace{0.5cm}Unique match & 1526 & 632 \\ 
     \hline
\end{tabular}
\end{center}
\end{table}
Evidently, the expected risk and the true match rate have been reduced (12 and 21 times, respectively) and the false match rate has increased significantly (from 0 to close to 90\%) with the synthesis process, suggesting a high level of identification disclosure risk reduction provided by our synthetic YRBS.\par

\subsubsection{Record linkage approach}

For partially synthetic data, record linkage methods can be applied to linking records in the synthetic dataset to the records in the confidential dataset. Based on variables, called keys, a link between two records can be established and we can evaluate identification risks in terms of true links and false links.\par



As with the matching-based approach, variables such as city, age, sex, grade, and race are considered as keys, i.e., the variables the intruder may use to establish the linkage, in our evaluation for the synthetic YRBS. For each record $i$ in the confidential YRBS, multiple linkages in the synthetic YRBS can be established, and the linkages are ranked by a weight estimated using the expectation-maximization algorithm by \citet{winkler_2000}. We use a greedy algorithm to search for the linkage with the highest weights for each record. The process of linkage establishment and greedy search are implemented by the \texttt{reclin} R package \citep{reclin}. 

Similar to the matching-based approach, we calculate the percentages of the true links and false links in both the synthetic and the confidential data for comparison. The confidential YRBS have a true linkage rate of $100\%$ and a false linkage rate of $0\%$ whereas the synthetic YRBS have a true linkage rate of $\mathbf{8.5\%}$ and a false linkage rate of $\mathbf{91.5\%}$. A 11-fold reduction in the true linkage percentage and a 0\% to 91.5\% increase in the false linkage percentage suggest that the synthetic YRBS make it much more difficult for an intruder to establish true record links based on the knowledge they possess, and therefore our synthesis process has successfully reduced identification disclosure risks significantly.\par




\subsection{Attribute Disclosure Risk}
To evaluate attribute disclosure risk, we consider two methods: the correct attribution probability (CAP) and the classification-based approach. The results show that our synthetic YRBS provide a significant attribute disclosure risk reduction compared to the confidential YRBS.\par

\subsubsection{Correct Attribution Probability (CAP)}
The CAP measures the probability that an intruder can correctly predict the value of the target variable for an individual by using the empirical distribution of this variable among synthetic observations with the same key variables. In our evaluation of the synthetic YRBS, the key variables are city, age, sex, grade, and race, and the target variable is marijuana usage. 

We follow the set-up in  \citet{baillargeon_and_charest}. Let $\mathbf{Y}$ denote the confidential dataset and $y_{ij}$ represents the $j$-th variable of the $i$-th record. For a specific sensitive variable $l$, all possible values for this variable are the targets denoted as $T_1,...,T_G$, where $G$ is the number of levels of the target variable. The intruder attempts to predict the value of $y_{il}$ using some or all of $Y^{-l}$, the set of variables other than $l$. These variables are the keys, denoted as $K_1,...,K_H$. The CAP of record $y_0$ in confidential dataset $\mathbf{Y}$ with synthetic dataset $\mathbf{Z}$ is given as:
\begin{equation}
    \text{CAP}_{y_0}(\mathbf{Z}) = \frac{\sum_{i=1}^n I[T(z_i)=T(y_0),K(z_i)=K(y_0)]}{\sum_{i=1}^nl[K(z_i)=K(y_0)]}.
    \label{eq:CAP}
\end{equation}
Equation (\ref{eq:CAP}) represents the proportion of target variable matches in all the key variable matches for a particular sensitive variable $l$ and a particular record $y_0$. The CAP for a synthetic dataset can be computed by averaging the CAP over all records. The average CAP for the $m = 5$ synthetic YRBS datasets is $\mathbf{0.749}$, while the CAP computed from the confidential dataset is $0.753$.\par

\begin{figure}[h!]
\begin{center}
\caption{Synthetic individual CAP versus confidential individual CAP given the marijuana variable.}\label{fig:individual_CAP}
    \includegraphics[width = 0.8\textwidth]{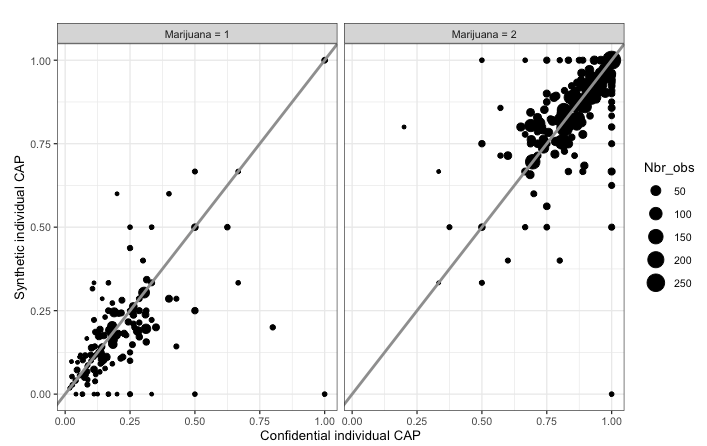}
\end{center}
\end{figure}

Comparing 0.749 and 0.753 indicates that the average CAP does not reduce much from the synthesis process for the file as a whole. However, it is important to note that the CAP for each record could be changed by the synthesis process to different extents which cannot be captured by the average CAP. To visualize the change of CAP at the individual level, Figure \ref{fig:individual_CAP} plots the synthetic individual CAP versus the confidential individual CAP by marijuana status. 



Figure \ref{fig:individual_CAP} shows that most records fall on the 45 degree line, meaning that for these records, there is no major difference in attribution probability before and after synthesis. It is also notable that most records with marijuana usage equals 1, i.e., they use marijuana, have a low CAP in both the confidential and synthetic data, indicating that these records are relatively safe and the true attribute value is hard to be inferred regardless of whether they are synthesized.\par

\subsubsection{Classification-based risk measure}
A weakness of the CAP measure is that is uses a simple model to predict the values of the target variable. With a classification model, sophisticated algorithms can be deployed to predict the value of the target variable using a set of keys. In our evaluation of the synthetic YRBS, we adopt a random forest classifier to perform the task of predicting the value of marijuana use, the same task in the CAP illustrative above. We use city, age, sex, grade, obesity, and sexuality as predictors.\par 

A random forest classifier contains a number of decision trees on various subsets of the given dataset and takes the average to improve the predictive accuracy of that dataset \citep{randomforest}. We use the synthetic data $\mathbf{Z}$ to train a model and test the model with the confidential data $\mathbf{Y}$ to evaluate the accuracy. In comparison, we also train a model using the confidential data and testing it on the confidential data. The algorithm is implemented by the \texttt{randomForest} package in R. 
The classification error for $\text{Marijuana}=1$ is $0.995$ on the confidential data and $\mathbf{0.988}$ on the synthetic data; the error for $\text{Marijuana}=2$ is $0$ on the confidential data and $\mathbf{0.004}$ on the synthetic data.



These results show that $\text{Marijuana} = 1$ is always difficult to predict: Even if we train the model with confidential data, the model performs poorly on capturing these data. In fact, the error rate has decreased if we train the model on the synthetic data, meaning that the risk is potentially higher in the synthetic data. However, since the marijuana variable is highly skewed, and the random forest classifier has a random component in it that every time it builds a slightly different model, we cannot conclusively state whether the synthesis process has increased or reduced the attribute disclosure risks.



In summary, two metrics of identification disclosure risks indicate that our synthetic YRBS have substantially reduced such risks, while the results are less conclusive for attribute disclosure risk evaluation. 

\section{Concluding remarks}
\label{sec:conclusion}
In conclusion, the respondent-level privacy and confidentiality in the YRBS data sample are well protected by the DPMPM synthesis model, especially in terms of identification disclosure risk reduction. At the same time, the synthetic YRBS preserve a high level of data utility, both in terms of global and analysis-specific utility with various metrics.


There are a few limitations of our case study. First, some of the utility and risk evaluation methods consider a few scenarios and some of the measurements such as the inference for a regression coefficient and the classification based approach are model-dependent. However, it is admittedly infeasible to comprehensively consider all possible inferences and prediction models that a data analyst or a intruder might use. Second, the YRBS data are highly skewed and unbalanced, especially in some of the sensitive variables. Such skewness can be challenging to capture by our synthesis model, as well as for a data analyst or a hypothetical intruder. We believe this is the main reason that for highly unbalanced variables the utility and the risk reduction results are not satisfactory. 

Despite these limitations, our case study serves as a useful demonstration of the DPMPM synthesis model and illustrates how widely useful and applicable it can be. The model can be extended to the rest of the YRBS survey (those before 2019 and other than NYC and Chicago), as well as some other categorical data in general. We also believe our case study showcases a variety of utility and disclosure risk evaluation metrics in practice, which can be useful and beneficial to data disseminators who are considering the synthetic data approach for microdata release. \par




\bibliographystyle{natbib}
\bibliography{reference.bib}

\appendix
\section{Density plots of differences in relative frequencies in Section 4.1}\label{density plots}



\begin{figure}[H]
    \centering
     \caption{Deviation plot for one-way table.}
    \includegraphics[width =0.9 \textwidth]{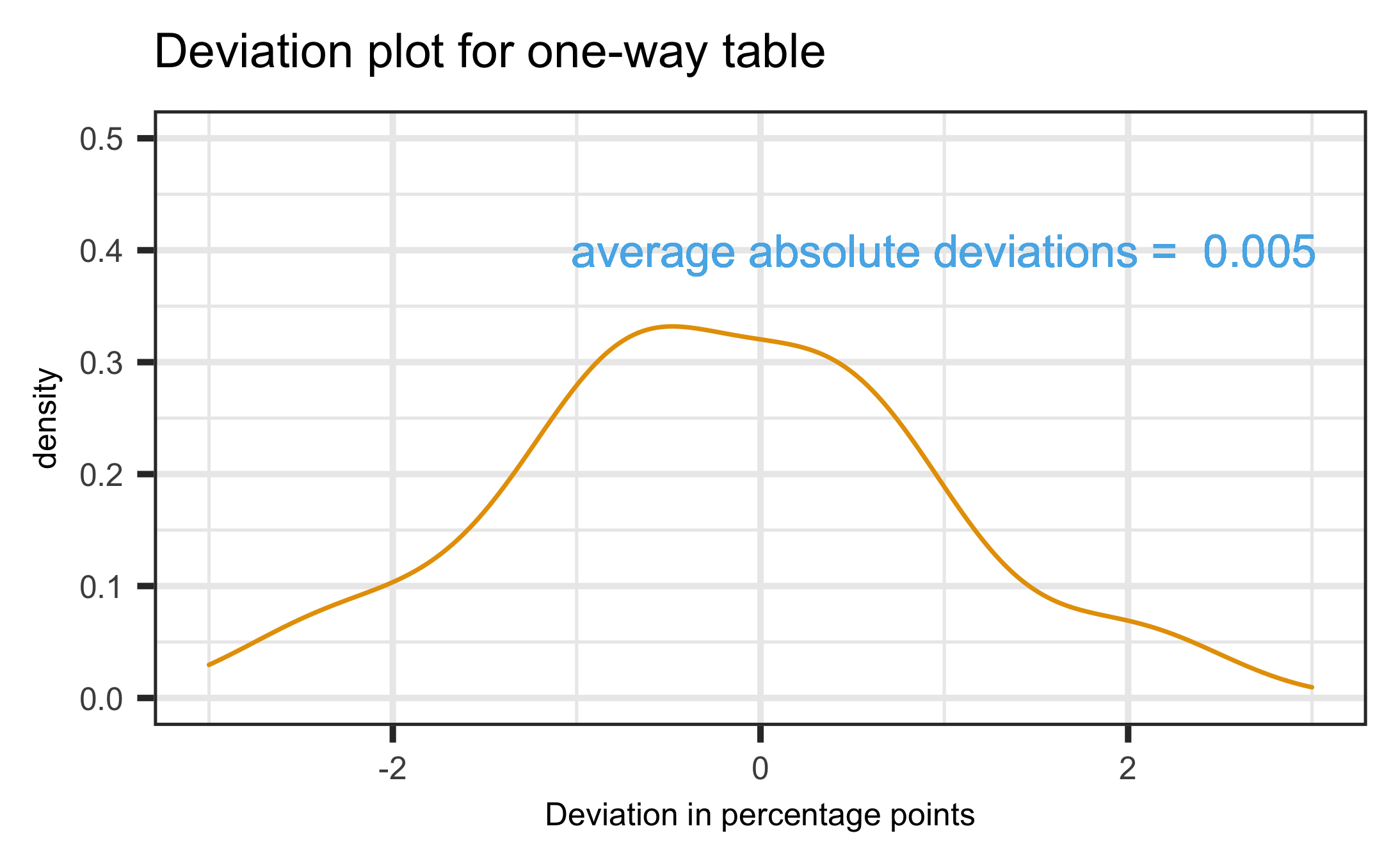}
    \label{fig:one_way_tab}
     \caption{Deviation plot for two-way table.}
    \includegraphics[width = 0.9\textwidth]{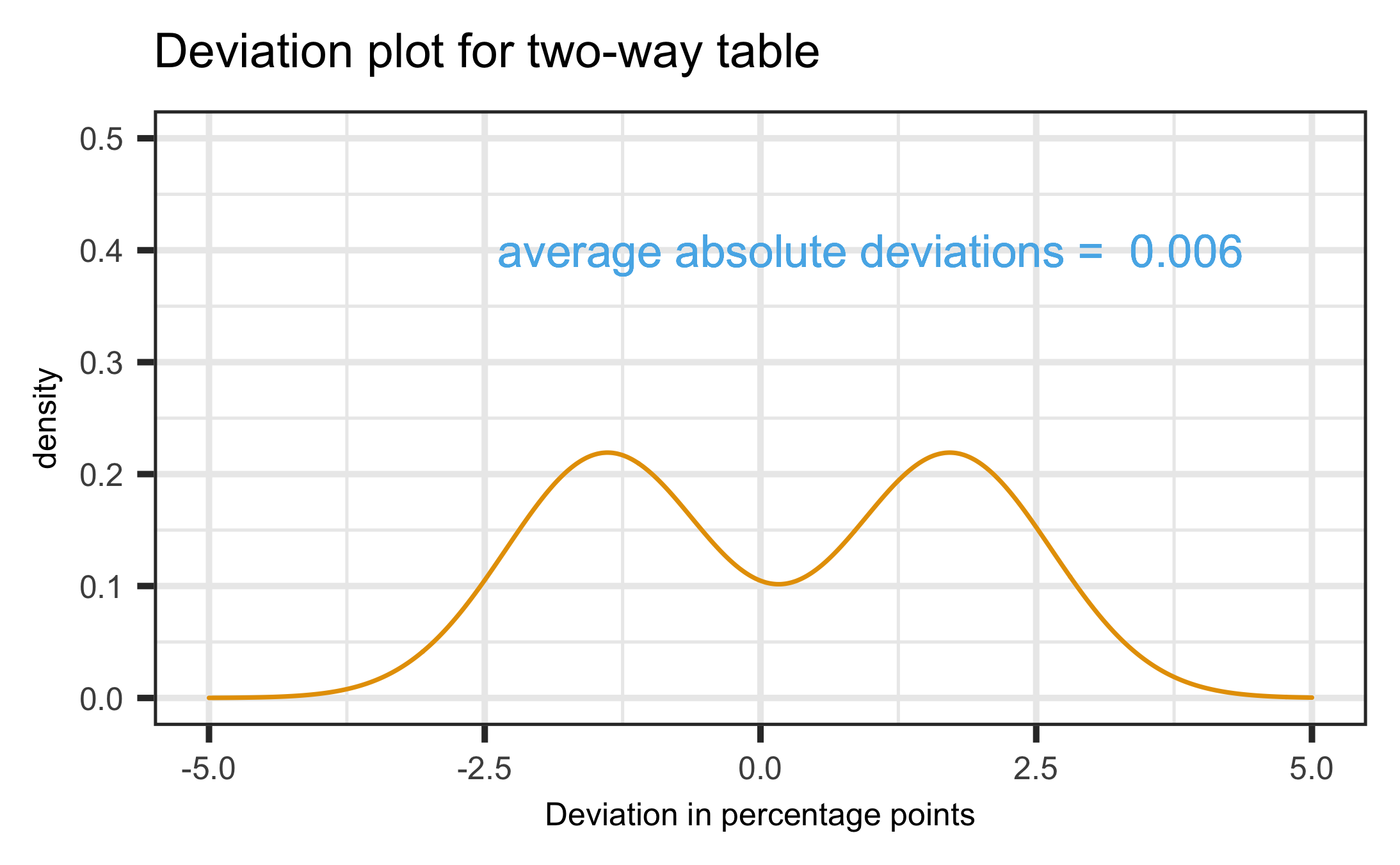}
    \label{fig:two_way_tab}
\end{figure}

\section{Three key quantities to evaluate disclosure risk in matching-based approach}\label{key_quantities}

The following set up is a basic version of \citet{drechsler_reiter_2009} to compute the three risk metrics.\par

We separate the vector of responses of the $i$-th record into two groups: variables available from external databases and variables unavailable to users except in the released data, denoted as $\mathbf{y}_i = (y_{i1},...y_{ir})=(\mathbf{y}_i^A, \mathbf{y}_i^U)$. We also have the matrix $\mathbf{Y} =(\mathbf{Y}^A,\mathbf{Y}^U)$ representing the confidential values of all $n$ units. On the confidential data holder side, similar to the split of $\mathbf{y}_i$, we have $\mathbf{z}_i = (z_{i1},...,z_{ir}) = (\mathbf{z}_i^A, \mathbf{z}_i^{U})$. We further split $\mathbf{z}_i^A$ into the synthesized variables $\mathbf{z}_i^{A_s}$ and the unsynthesized variables $\mathbf{z}_i^{A_{us}}$ , and let $\mathbf{Z} = (\mathbf{Z}^{A_{us}}, \mathbf{Z}^{A_s}, \mathbf{Z}^U)$ be the matrix of all released data. On the intruder side, let $\mathbf{t}$ be the vector of information available to the intruder, we assume $\mathbf{t}=\mathbf{y}^A$ for some unit in the population, that is, the intruder obtains their knowledge about the dataset from some external database. Additionally, let $S$ denote the meta-data released about the simulation models used to generate the synthetic data and $R$ denote the meta-data released about the reason why records were selected for synthesis. In our basic version, we assume $S$ and $R$ to be both empty. Let $l$ be the random variable that equals $i$ when $z_{i0}=t_0$ for $i\in\mathbf{Z}$ and equals to $n+1$ when $z_{i0}=t_0$ for $i\notin \mathbf{Z}$, where index $0$ denotes the ``zero column'', a unique ID for each record. Then the intruder is interested in calculating for $i=1,...,n+1$
$$Pr(l=i|\mathbf{t}, \mathbf{Z}, S, R)$$
The three risk summaries can then be computed as follows:
\begin{align}
    \text{Expected Match Risk} &= \sum_{i=1}^n\frac{T_i}{c_i} \\
    \text{True Match Rate} &= \sum_{i=1}^n\frac{K_i}{N} \\
    \text{False Match Rate} &= \sum_{i=1}^n\frac{F_i}{s}
\end{align}
where $c_i$ is the number of records with the highest match probability for record $i$, $T_i=1$ if the true match is among the $c_i$ units and $T_i=0$ otherwise, $K_i=1$ if the true match is the unique match and $K_i=0$ otherwise, $N$ is the total number of target records out of $n$ records, $F_i=1$ if there is a unique match but it is not the true match and $F_i=0$ otherwise, $s$ is the number of unique matches. 
\end{document}